\documentclass[twocolumn,aps, prl, groupedaddress,epsf]{revtex4}

\usepackage{graphicx}

\begin{document}
%\draft \preprint{July 12, 2007}

\title{Measurement of Scattering Rate and Minimum Conductivity in Graphene}

\author{Y.-W. Tan$^{1\dag}$}
\author{Y. Zhang$^{1\dag}$}
\author{K. Bolotin$^1$}
\author{Y. Zhao$^1$}
\author{S. Adam$^2$}
\author{E.~H.~Hwang$^2$}
\author{S. Das Sarma$^2$}
\author{H. L. Stormer$^{1, 3, 4}$}
\author{P. Kim$^1$}

\affiliation{$^1$Department of Physics, Columbia University, New
York, New York 10027}

\affiliation{$^2$Condensed Matter Theory Center, Department of
Physics, University of Maryland, College Park, Maryland 20742}

\affiliation{$^3$Department of Applied Physics, Columbia
University, New York, New York 10027}

\affiliation{$^4$Bell Labs, Alcatel-Lucent Technologies, Murray
Hill, NJ 07974}

\begin{abstract}
The conductivity of graphene samples with various levels of
disorder is investigated for a set of specimens with mobility in
the range of $1-20\times10^3$~cm$^2$/V~sec.  Comparing the
experimental data with the theoretical transport calculations
based on charged impurity scattering, we estimate that the
impurity concentration in the samples varies from $2-15\times
10^{11}$~cm$^{-2}$. In the low carrier density limit, the
conductivity exhibits values in the range of $2-12e^2/h$, which
can be related to the residual density induced by the
inhomogeneous charge distribution in the samples. The shape of the
conductivity curves indicates that high mobility samples contain
some short range disorder whereas low mobility samples are
dominated by long range scatterers.
\end{abstract}


\maketitle

%\ START OF TEXT BODY

The recent discovery of graphene~\cite{Novoselov04Sci}, a single
atomic sheet of graphite, has created an intense research effort
on this new material. The uniqueness of the electronic band
structure of graphene is exemplified by the charge neutral Dirac
points located at the Brillouin zone corner. The linear dispersion
relation of the bands crossing these Dirac point can be
interpreted in analogy to relativistic fermions of vanishing mass.
The nontrivial Berry phase associated with the graphene band
structure accounts for unusual half integer shifted quantum Hall
effect ~\cite{Novoselov05Nat, Zhang05Nat} in the presence of a
magnetic field. This chiral nature of the Dirac fermions is also
expected to induce nontrivial charge transport phenomena, such as
a minimum conductivity ~\cite{Novoselov05Nat,Geim07NatM,
Nilsson06, Katsnelson06, Nomura06} and weak
anti-localization~\cite{Suzuura02, Khveshchenko06, McCann06,
Aleiner06}. Indeed, transport measurements indicate the existence
of such a minimum conductivity ~\cite{Novoselov05Nat,Geim07NatM}
and the suppression of localization~\cite{Morozov06} in some
graphene samples. Despite qualitative agreement, however, these
experimental observations do not agree quantitatively with the
theoretical expectations~\cite{Novoselov05Nat, Zhang05Nat,
Geim07NatM}. In order to understand these and other low energy
transport phenomena in graphene, it is essential to study relevant
carrier scattering mechanisms.

In this Letter, we report low magnetic field transport
measurements on 19 graphene devices with various levels of
disorder. We infer the scattering mechanisms using the density
dependence of the mean free paths and phase coherence lengths and
comparing with the corresponding transport theory.  The sample
mean free path is extracted from the conductivity measurements,
and we find that at high carrier density, two different scattering
mechanisms determine the density-dependent conductivity.

\begin{figure}
\includegraphics[width=75mm]{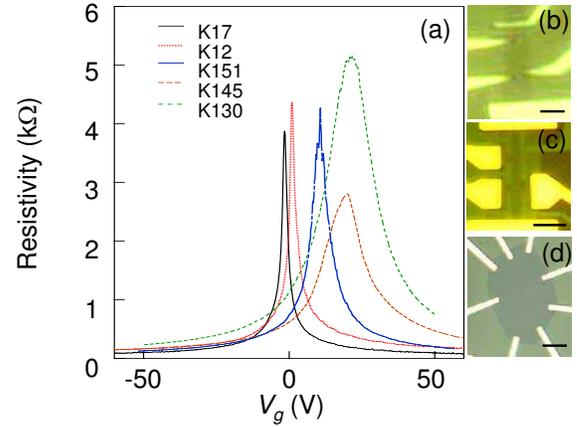}
\caption{(color online) (a) Resistivity of four representative
graphene samples as a function of applied gate voltage. All data
were measured at temperature 1.6~K. (b-d) Optical microscope
images of typical devices. The scale bar represents, 1, and
10~$\mu$m, for (b) and (c, d), respectively} \label{fig1}
\end{figure}

The graphene pieces used in this work are extracted from bulk
graphite crystals and deposited onto SiO$_2$/Si substrate using
the mechanical method described in previous
work~\cite{Novoselov05PNAS}. Typically, graphene pieces of lateral
size 3-20~$\mu$m are chosen for device fabrication. For a few
large flake samples (lateral size $>20~\mu$m), we explicitly
fabricated well-defined two to three Hall-bar geometry devices
having the same width but varying lengths. These multiple devices
from the same graphene piece provide means to check the
consistency of our conductivity measurements. The metallic
electrodes are defined on the sample using electron beam
lithography followed by Au/Cr (30/3~nm) evaporation and a lift-off
process. The degenerately doped silicon substrate serves as a gate
electrode with 300~nm thermally grown silicon oxide acting as the
gate dielectric. By applying a gate bias voltage, $V_g$, the
charge density of the sample can be tuned. The conductivity of a
sample is measured either in the Hall bar geometry considering the
width and length of the samples in between the voltage probes
(Fig.~1(b)) or by van der Pauw method for samples with irregular
shape (Fig.~1(c)). We also use Hall-bar shaped devices (Fig.~1(d))
in order to minimize systematic measurement errors.  All samples
are measured using lock-in amplifiers at an excitation current
less than 50~nA to minimize any heating effect on the samples,
which are kept in the helium vapor of a cryostat.

Fig.~1(a) shows the resistivities, $\rho$, of five representative
samples as a function of $V_g$. The resistivity curves are largely
symmetric around a particular gate voltage $V_g=V_{Dirac}$ and
show a maximum at this value.  The finite value of $V_{Dirac}$
indicates that there exists an unintentional doping of the
graphene samples~\cite{Hwang07} whose origin may be electrostatic
or caused by charged impurities. The actual carrier density $n$ in
graphene induced by the gate voltage in the presence of impurity
doping is then obtained from $n=C_g(V_g -V_{Dirac})/e$ where the
gate capacitance $C_g=115$~aF/$\mu$m$^2$ is deduced from a
separate Hall measurement. We observe that $\rho$ decreases
monotonically as $|n|$ increases, however the sharpness of the
dips at $n=0$ varies drastically from sample to sample.

\begin{figure}
\includegraphics[width=80mm]{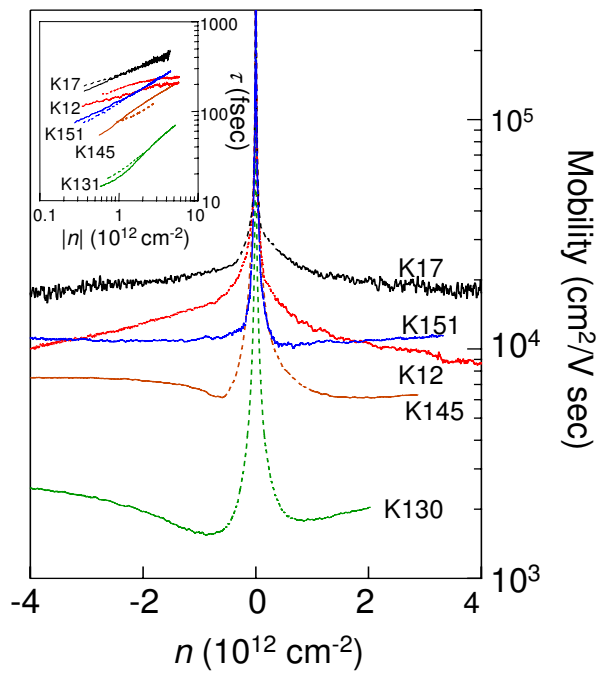}
\caption{(color online) Mobility estimated by applying Drude model
to the data in Fig.~1. The graphs shown by broken lines are the
region where simple Drude model fails to hold. The inset shows the
scattering time in this model. The solid lines are for electrons
and dashed lines are for holes.} \label{fig2}
\end{figure}

In order to analyze the difference between samples quantitatively,
we employ the semi-classical Drude model to estimate the mobility
of samples, $\mu=(en\rho)^{-1}$. Fig.~2 display $\mu$ as a
function of $n$ obtained from Fig.~1(a). Although $\mu$ defined in
this way diverges at $V_g=V_{Dirac}$ (the portion of curves
represented by broken lines) due to the trivial reason that $n=0$
nominally at the Dirac point, the limiting value of the mobility,
$\mu_{L}$, defined in the large density limit ($n\sim4 \times
10^{12}$~cm$^{-2}$) serves as a useful way to characterize the
sample quality. We have measured a total of 19 samples in this
experiment where $\mu_{L}$ ranges $2,000-20,000$~cm$^2$/Vsec.
Quite generally, for samples of poorer quality, the Dirac points
shift more from $V_g=0$, indicating a larger, unintentional charge
doping of unknown origin~\cite{Wehling07}. The inset to Fig.~2
shows explicitly the scattering time $\tau$ of the samples shown
in Fig.~1, with $\tau = \mu E_F/(e v_F^2)$ by definition.  The
corresponding mean free path of the samples can be obtained from
$l=v_F\tau$, where the Fermi velocity $v_F\approx
10^6$~m/sec~\cite{Geim07NatM}. We note that the resulting mean
free path for both electrons and holes strongly depends on the
density $n$, ranging from 10-500~nm in the most of the samples and
density ranges.

The quality of the samples can be further differentiated by comparing
their behavior of carrier density dependent conductivity. Fig.~3 shows
$\sigma=\rho^{-1}$ as a function of $V_g$. Evidently, the samples with
poorer quality ($\mu_{L}<5,000$~cm$^2$/Vsec) show a very broad and
smooth maximum in $\rho(V_g)$ near the Dirac point followed by a
linear relationship of $\sigma(V_g)$ in the large density limit. On
the other hand, when the mobility of a sample is larger than
$\sim10,000$~cm$^2$/Vs, $\sigma(V_g)$ forms a cusp around the Dirac
point followed by a sublinear increase both in the electron and hole
regimes. Since the scattering time $\tau$ goes as $\tau\sim \sigma
n^{-1/2}$, this difference in the density dependence of $\sigma$ in
the two different mobility regimes strongly indicates that different
scattering mechanisms might be dominating the high carrier density
transport in samples from the two limiting groups.  Recent numerical
calculations based on Ref.~\cite{Nomura06,Hwang07} indeed show that
$\sigma(n)$ changes from a linear dependence to a sublinear dependence
as the scattering mechanism changes to short range scattering, such as
atomic defects in the lattice, from long range scattering, such as
ionized impurity scattering.

\begin{figure}
\includegraphics[width=75mm]{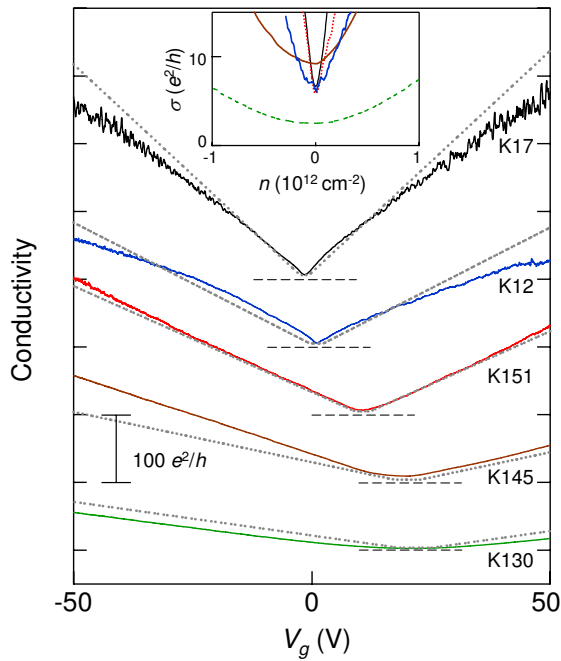}
\caption{(color online)The conductivity of five representative
samples shown in Fig.~1 as a function of applied gate voltage
(solid lines). For clarity, the each curves are displaced. The
horizontal dashed lines indicate the zero conductance for each
curves. Dotted curves are the corresponding theoretical results as
explained in the text. The only fit parameter in the theory is the
density of the random charged impurity centers in the substrate,
which is taken to be (top to bottom in units of
10$^{11}$~cm$^{-2}$): 2.2, 4.0, 4.6, 9.7, 14.5. The inset shows
the detailed view of the density dependent conductivity near the
Dirac point for the data in the main panel.} \label{fig3}
\end{figure}

Comparing our experimental data in a quantitative fashion with
these theoretical calculations~\cite{Hwang07}, we estimate that
the charged impurity concentration varies in the range of
$2-15\times10^{11}$~cm$^{-2}$ . Compared to recent Raman data on
graphene samples, where the typical impurity density $n_i$ is
estimated to be $n_i\sim 3\times10^{11}$~cm$^{-2}$ , this value
appears to be quite reasonable~\cite{Yan07}.  We note that the
relevant scattering times entering Raman scattering level
broadening and dc conductivity are not necessarily identical, and
typically the level broadening is larger than the transport
broadening for the same impurity density. This explains why the
Raman impurity density estimate falls toward the lower end of our
transport estimate.

Simple theoretical arguments show that long ranged charged
impurity scattering is much more effective for scattering of
carriers than short-range scattering~\cite{Nomura06,Hwang07}.  For
example, given the same impurity density $n_i \sim
5\times10^{11}$~cm$^{-2}$, $l\gtrsim 1~\mu$m for short-range
scatterers but $l\sim 50$~nm for charged impurity scatterers.
Since the size of our samples $L$ is on the order of several
microns, transport can be considered to be in the diffusive
regime.  A Boltzmann transport theory with charged impurity
scattering can then be applied to solve for the conductivity
self-consistently, and we obtain~\cite{Adam07}
\begin{equation}\label{eq1}
\sigma(V_g)\approx\left\{ \;\;
  \begin{array}{ll}
  C\frac{e^2}{h}\frac{n}{n_i}& \mbox{for $n>n^*$}\\\\
  C\frac{e^2}{h}\frac{n^*}{n_i} & \mbox{for $n<n^*$}
  \end{array}
 \right.
\end{equation}
where the carrier density is related to the gate voltage by
$n=C_gV_g/e+\bar{n}$; $n^*$ describes the self-consistent residual
carrier density considering electron and hole puddle formation
induced by the charged impurities; $\bar{n}$ is the induced
carrier density in graphene by the charged impurity; and $C$ is a
dimensionless numerical parameter describing the strength of
scattering by the potential fluctuation considering full random
phase approximation (RPA) screening. Considering the dielectric
screening from the SiO$_2$ substrate underneath the graphene
samples, we estimate $C\approx 20$~\cite{Adam07}. Within the RPA,
the values of $n^*$ and $\bar{n}$ can be self-consistently
determined by the charged impurity concentration $n_i$, and $d$,
the distance between the graphene sheet and the charged impurity
layer~\cite{Galitski07,Adam07}.

We emphasize that in this model with a single free parameter
$n_i$, Eq.~\ref{eq1} describes: (i) the linear tail mobility, (ii)
the offset voltage $V_{Dirac}=-e\bar{n}/C_g$ , (iii) the minimum
conductivity given by $\approx 20e^2/h (n^*/n_i)$ and (iv) the
plateau width i.e. the range in gate voltage over which the
conductivity saturates at its minimum value. Since these features
are determined by the impurity concentration only, we now
understand qualitatively that the behavior of conductivity of the
cleaner samples (i.e. those with lower charged impurity density),
in addition to having a higher slope, also have narrower minimum
conductivity plateaus, smaller values of $V_{Dirac}$, and higher
values of the minimum conductivity (i.e.,
$\sigma(V_g=V_{Dirac})$). On the other hand, the dirtier samples
have lower mobilities, wider plateau widths, and larger gate
voltage offsets.

For a quantitative comparison, the theoretical model
(Eq.~\ref{eq1}) is now compared to the experimental data (Fig.~3).
To avoid having too many fitting parameters in the theory, we just
set $d\approx1$~nm (which is typical for SiO$_2$
substrate~\cite{Adam07}), keeping only $n_i$ as a single fitting
parameter. This yields corresponding charged impurity
concentrations for the samples in Fig.~3 ranging between
$2-15\times 10^{11}$~cm$^{-2}$, where a reasonable agreement
between experiment (solid curves) and theoretical curves can be
seen. We note that that the some discrepancies between the
experimental data and the simple model can be further understood:
for example the sub-linear behavior seen in high quality samples
(K17 and K12) is consistent with short-range scattering that
becomes important only at very high density~\cite{Hwang07}. The
electron and hole asymmetry often observed in poor quality samples
(K130) can be explained by drift of charged impurity in the
substrate. The large gate voltages used in the experiment
($V_g\sim50$~V) can shift the impurity trapped in the substrate,
resulting in the adjustment of $d$ for electrons and holes. Had we
introduced two additional fit parameters (one to parameterize the
amount of short-range scattering and another one to parameterize
$d$ the distance between the charged impurities and the graphene
sheet) we would have obtained exact quantitative agreement. Given
the reasonable agreement between theory and experiments for
several different samples in Fig.~3, we do not see any advantages
in obtaining an exact quantitative agreement by doing this
additional data fitting.

\begin{figure}
\includegraphics[width=75mm]{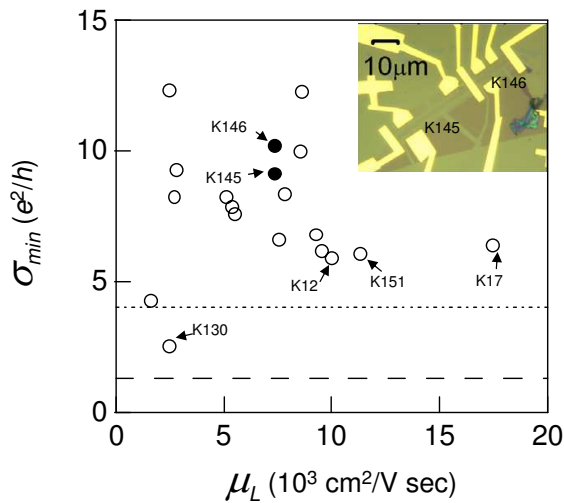}
\caption{The minimum conductivity and limiting mobility values at
metallic regime for all 19 samples measured at 1.6~K. Filled
symbols indicate two devices, K145 and K146 made out of the same
graphene flake as shown in the optical microscope image in the
inset.} \label{fig4}
\end{figure}

We now address the minimum value of the conductivity,
$\sigma_{min}$ observed in our samples. Fig.~4 displays
$\sigma_{min}$ versus the high-density, limiting mobility value
$\mu_L$ for all our samples. The measured values of $\sigma_{min}$
lie in the range of $2-12e^2/h$. Most samples have
$\sigma_{min}>4e^2/h$, in contrast to an earlier
study~\cite{Novoselov05Nat} reporting a clustering of
$\sigma_{min} \approx4e^2/h$ (horizontal dotted line).  Several
theories~\cite{Nilsson06, Nomura06} based on homogeneous graphene
sheets predict a universal value of $4e^2/\pi h$ for
$\sigma_{min}$ (horizontal dashed line), which is almost an order
of magnitude lower than in most of our samples. The sample
dependent $\sigma_{min}$ was recently argued to arise from the
inhomogeneous charge distribution near the Dirac
point~\cite{Geim07NatM, Hwang07}. More recently, theoretical work
based on numerical calculation also indicates a non-invariant
$\sigma_{min}$ scaling with the system size in clean limit
~\cite{Bardarson07, Nomura07}. Further careful consideration is
needed for disordered samples. In the low density limit near the
Dirac point where the carrier concentration becomes smaller than
the charged impurity density, the system breaks up into puddles of
electrons and holes where a duality in two dimensions guarantees
that locally transport occurs either through the hole channel or
the electron channel. Percolation resistance of these electron and
hole puddles are the key to understanding the experimentally
observed minimum conductivity~\cite{Falko07}. The charged
impurities induce a density distribution in the graphene sample
where the presence of both electrons and holes carriers implies
that both positive and negative fluctuations are screened by
graphene~\cite{Adam07}. It is the residual density given by $n^*$
that is responsible for the minimum conductivity, and therefore we
expect that the value of the minimum conductivity will depend on
the impurity concentration. Very recent scanning single electron
transistor microscopy demonstrated such electron-hole puddle
formation near the charge neutral point in
graphene~\cite{Martin07}.

In conclusion, we report zero-field transport properties of more
than a dozen graphene samples with different mobility values. The
minimum conductivity is found to be strongly sample dependent,
yielding values an order of $e^2/h$ with a non-universal
prefactor. Density inhomogeneities across the graphene sheet may
be responsible for this large spread in conductivity.

We thank A. Geim, J. Yan, and I. Aleiner for helpful discussions.
This work is supported by the DOE (DE-FG02-05ER46215 and
DMR-03-52738), ONR (N000150610138), NSEC (CHE-06-41523 and
DMR-03-52738), W. M. Keck foundation and the New York State Office
of Science, Technology, and Academic Research (NYSTAR). The work
at Maryland is supported by US-ONR, LPS-NSA, and Microsoft
Q-Project.

$^\dag$ Current address: Dept. of Chemistry/Physics, University of
California at Berkeley.


\begin{references}

\bibitem{Novoselov04Sci} K. S. Novoselov,  A. K. Geim, S. V.
Morozov, D. Jiang, Y. Zhang, S. V. Dubonos, I. V. Grigorieva, A.
A. Firsov, Science {\bf 306}, 666 (2004).

\bibitem{Novoselov05Nat} K. S. Novoselov, A. K. Geim, S. V.
Morozov, D. Jiang, M. I. Katsnelson, I. V. Grigorieva, S. V.
Dubonos, A. A. Firsov, Nature {\bf 438}, 197 (2005).

\bibitem{Zhang05Nat} Y. Zhang, Y. -W. Tan, H. L. Stormer,
P. Kim, Nature {\bf 438}, 201 (2005).

\bibitem{Geim07NatM} A. K. Geim, K. S. Novoselov,
Nature Mat. {\bf 6}, 183 (2007).

\bibitem{Nilsson06} J. A. Nilsson, A. H. Castro Neto,
F. Guinea, and N. M. R. Peres, Phys. Rev. Lett. {\bf 97}, 266801
(2006).

\bibitem{Katsnelson06} M. I. Katsnelson, Eur. Phys. J. B {\bf 51},
157-160 (2006).

\bibitem{Nomura06} K. Nomura, and A. H. MacDonald, Phys. Rev. Lett.
{\bf 98}, 076602 (2007)

\bibitem{Suzuura02} H. Suzuura, and T. Ando, Phys. Rev. Lett.
{\bf 89} 266603 (2002).

\bibitem{Khveshchenko06} D. V. Khveshchenko, Phys. Rev. Lett. {\bf 97}
036802 (2006).

\bibitem{McCann06} E. McCann, K. Kechedzhi, V. I. Fal'ko, H. Suzuura, T. Ando,
B. L. Altshuler, Phys. Rev. Lett. {\bf 97} 146805 (2006).

\bibitem{Aleiner06} I. L. Aleiner, and K. B. Efetov, Phys. Rev. Lett.
{\bf 97} 236801 (2006)

\bibitem{Morozov06} S. V. Morozov,  K. S. Novoselov, M. I. Katsnelson,
F. Schedin, L. A. Ponomarenko, D. Jiang, A. K. Geim, Phys. Rev.
Lett. {\bf 97} 016801 (2006).

\bibitem{Novoselov05PNAS} K. S. Novoselov, D. Jiang, T. Booth, V. V.
Khotkevich, S. M. Morozov, A. K. Geim, Proc. Natl. Acad. Sci. U.S.A.
{\bf 102}, 10451 (2005).

\bibitem{Hwang07} E. H. Hwang, S. Adam, S. Das Sarma, Phys. Rev. Lett.
{\bf 98} 186806 (2007);

\bibitem{Wehling07} T. O. Wehling, K. S. Novoselov, S. V. Morozov,
E. E. Vdovin, M. I. Katsnelson, A. K. Geim, A. I. Lichtenstein,
arXiv:cond-mat: 0703390 (2007);

\bibitem{Yan07} J. Yan, Y. Zhang, P. Kim, and A. Pinczuk,
Phys. Rev. Lett. {\bf 98} 166802 (2007).

\bibitem{Adam07} S. Adam, E. H. Hwang, V. M. Galitski, S. Das Sarma,
arXiv:cond-mat: 0705.1540 (2007).

\bibitem{Galitski07} V. M. Galitski, S. Adam and S. Das Sarma,
arXiv:cond-mat: 0702117 (2007);

\bibitem{Bardarson07} J. H. Bardarson, J. Tworzydlo, P. W. Brouwer,
and C. W. J. Beenakker, arXiv:cond-mat: 07050886 (2007).

\bibitem{Nomura07} K. Nomura, M. Koshino, and S. Ryu, arXiv:0705.1607.

\bibitem{Falko07} V. V. Cheianov, V. I. Falko, B. L. Altshuler, I. L. Aleiner,
arXiv:cond-mat: 07062968 (2007).

\bibitem{Martin07} J. Martin, N. Akerman, G. Ulbricht, T. Lohmann,
J. H. Smet, K. von Klitzhing, A. Yacoby, arXiv:cond-mat: 07052180 (2007).

\end{references}
\end{document}